\documentclass[11pt]{article}

\usepackage[final]{acl}

\usepackage{times}
\usepackage{latexsym}

\usepackage[T1]{fontenc}

\usepackage[utf8]{inputenc}

\usepackage{microtype}
\usepackage{amsmath}
\usepackage{amssymb}
\usepackage{enumitem}
\usepackage{hyperref}
\usepackage{url}
\usepackage{booktabs}
\usepackage{amsfonts}
\usepackage{subfig}
\usepackage{xcolor}
\usepackage{inconsolata}
\usepackage[table]{xcolor}
\usepackage{booktabs}
\usepackage{multirow}
\usepackage{tabularx}
\usepackage{fontawesome5}
\usepackage{hyperref}
\newcommand{\redbg}[1]{\cellcolor{red!#1}}
\newcommand{\bluebg}[1]{\cellcolor{blue!#1}}
\usepackage{graphicx}

\definecolor{softred}{RGB}{205,92,92}
\definecolor{softgreen}{RGB}{76,153,102}

\newcommand{\relup}[1]{  {\scriptsize\textcolor{softgreen}{\,(+#1\%)}}}
\newcommand{\reldown}[1]{  {\scriptsize\textcolor{softred}{\,(-#1\%)}}}
\newcommand{\rw}[1]{\textcolor{black}{#1}}

\title{Aligned but Flattened: Analyzing the Trade-off between Cultural Alignment and Diversity in LLMs}

\author{
  {Jingshen Zhang}$^{1,2*}$,
  {Shaoyang Xu}$^{2}$,
  {Wenxuan Zhang}$^{2,\dagger}$
  \\
  \normalfont\textsuperscript{1}School of Computer Science and Technology, Tianjin University
  \\
  \normalfont\textsuperscript{2}iNLP Lab, Singapore University of Technology and Design
  \\
  \texttt{jason\_zhang@tju.edu.cn, wxzhang@sutd.edu.sg} \\ \\
  \faGlobe\hspace{0.3em}\url{https://aligned-but-flattened.github.io/}
}

\begin{document}
\maketitle

\begingroup
\renewcommand{\thefootnote}{\fnsymbol{footnote}}
\footnotetext[1]{ Work done during internship at the iNLP Lab}
\footnotetext[2]{ Corresponding author}
\endgroup

\begin{abstract}
Cultural fine-tuning has become the de facto paradigm for building culture-aware large language models (LLMs), yet existing optimization exclusively for alignment scores provides an incomplete portrait of cultural fidelity by systematically obscuring inherent cultural diversity. This unidimensional evaluation lens prompts a fundamental question: do models genuinely perceive distinct cultural nuances, or do they merely memorize dominant cultural values? To address this, we propose a synergistic evaluation framework that jointly formalizes cultural alignment and diversity. Through extensive benchmarking of six mainstream LLMs on the World Values Survey, this framework uncovers a systematic and critical trade-off: the pursuit of cultural alignment consistently incurs an acute expense of diversity, leading to severe "cultural flattening." Investigating this behavioral shift, we demonstrate that these superficial alignment gains stem from models artificially anchoring to dominant majorities, converging onto a monolithic response pattern that wipes out the heterogeneous distributions inherent to human groups. Crucially, our mechanistic analysis suggests that this diversity collapse is not merely a behavioral anomaly but more likely a structural consequence of the low-rank bias inherent in neural network optimization. Therefore, our findings expose the limitations of current post-training paradigms and call for a shift toward alignment objectives that preserve cross-cultural pluralism.
\end{abstract}

\section{Introduction}
\begin{figure}[t]
  \includegraphics[width=\columnwidth]{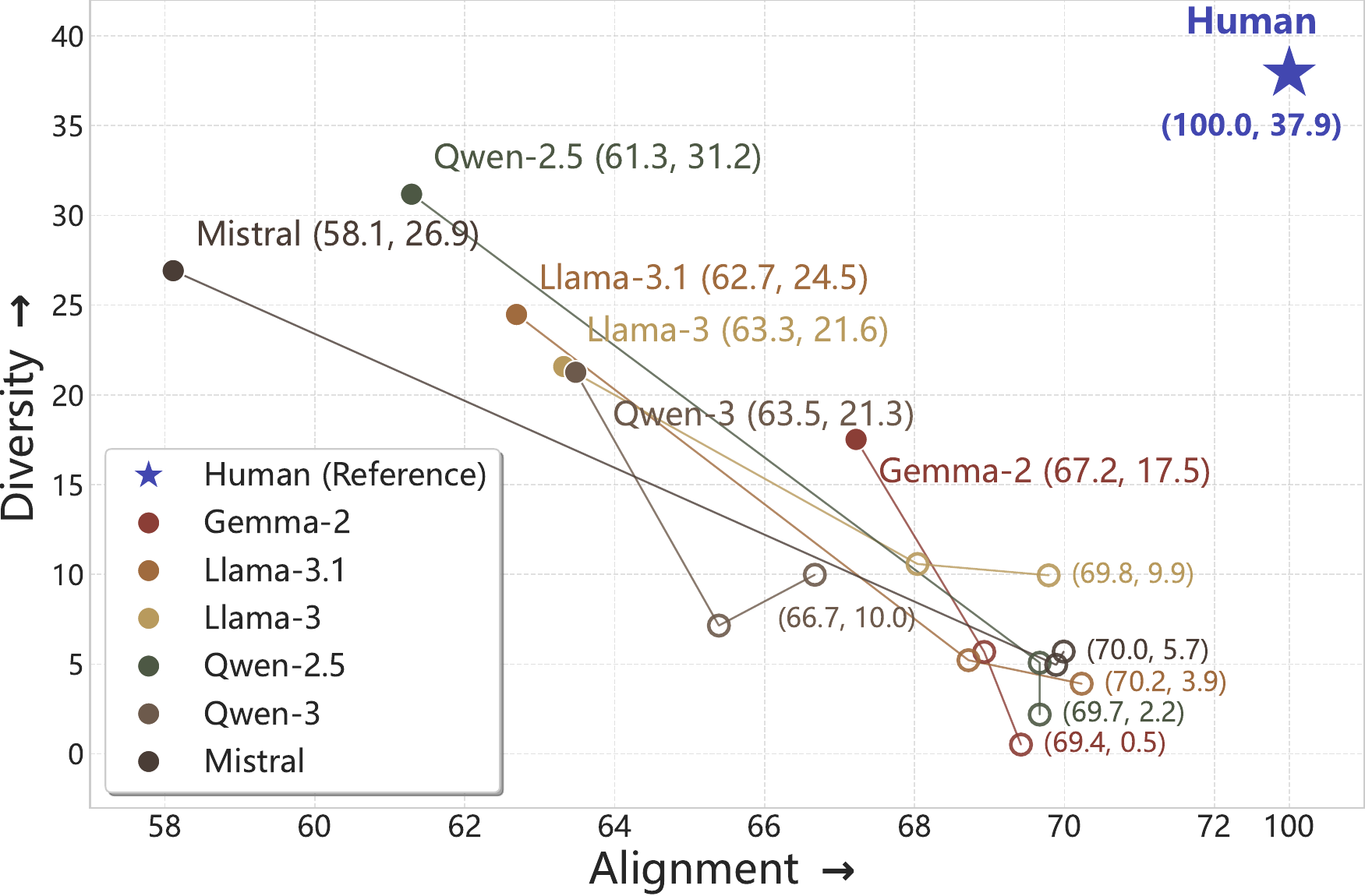}
  \caption{Trade-off between Cultural Alignment and Diversity. \textit{Solid} dots represent models \textit{without fine-tuning}, while \textit{hollow} circles represent \textit{fine-tuned} ones. For each model, different configurations are connected sequentially in descending order of cultural alignment.
  }
  \label{fig:trade-off}
\end{figure}

\rw{Given the rapid progress and the capacity to capture human behavioral patterns \citep{openai2024gpt4technicalreport, geminiteam2025geminifamilyhighlycapable, deepseekai2025deepseekv3technicalreport, yang2025qwen3technicalreport}, large language models (LLMs) are increasingly deployed worldwide as cultural agents, used to simulate the values and opinions of specific sociodemographic groups \citep{liu-etal-2025-cultural, kolluri-etal-2025-finetuning, hu2026simbenchbenchmarkingabilitylarge}.
For such uses, a model's \textit{\textbf{cultural alignment}}, i.e., how faithfully it reproduces the values of the population it represents, becomes essential. To improve it, a growing body of work fine-tunes LLMs on culture-specific data and consistently reports substantial gains in alignment scores \citep{NEURIPS2024_9a16935b, li2024culturepark, Xu2024SelfPluralisingCA, jiang-etal-2025-language}, making cultural fine-tuning the de facto recipe for building culture-aware models.}

\rw{However, a high alignment score alone provides an incomplete picture of cultural fidelity. Since alignment is usually optimized and reported as an aggregate similarity score, a model can improve alignment by tracking dominant response patterns while failing to preserve the diversity of opinions within and across cultures. This raises a central concern that remains underexplored in existing fine-tuning studies: \textit{does the pursuit of cultural \textbf{alignment} come at the expense of cultural \textbf{diversity}?} Our findings reveal that cultural fine-tuning consistently shifts models toward higher alignment, but also sharply reduces their behavioral diversity, revealing a systematic trade-off between the two, as shown in Figure~\ref{fig:trade-off}.}

\rw{
Unlike prior studies that primarily audit diversity in out-of-the-box models post hoc \citep{wu2024generativemonoculturelargelanguage, murthy-etal-2025-one}, we examine the cultural fine-tuning process itself---the stage at which this trade-off emerges and can therefore be analyzed and potentially mitigated. We first formalize cultural alignment and cultural diversity as two complementary evaluation axes. Alignment measures whether model responses match the corresponding human sociodemographic group, while diversity measures whether responses remain sensitive to differences across cultural personas. This formulation allows us to jointly evaluate whether a model becomes more culturally accurate while still preserving meaningful cross-cultural variation (§\ref{sec:preliminaries}).
}

\rw{
Using this framework, we fine-tune six LLMs on culture-specific data derived from the World Values Survey and benchmark their behavioral shifts against real human responses. Across all models, fine-tuning improves cultural alignment, but consistently narrows behavioral diversity. Importantly, this reduction is not merely a uniform decrease in variance; instead, it takes two distinct forms depending on the question type. For categorical questions, fine-tuned models collapse toward dominant cultural response modes, reducing a broad distribution of human opinions to one or two highly concentrated answers. For ordinal questions, models are susceptible to retreating toward a conservative middle option that covers nearby responses but fails to reproduce the full human distribution. In both cases, personas that should produce culturally differentiated responses are pushed toward more homogeneous outputs, creating the risk that non-dominant cultural perspectives are overwritten (§\ref{sec:experiment-on-cultural-alignment}).
}

\rw{
We further investigate this trade-off from a mechanistic perspective. Building on the low-rank simplicity bias of neural network optimization \citep{huh2023lowranksimplicitybiasdeep, súkeník2024neuralcollapseversuslowrank, galanti2024sgdweightdecaysecretly}, we hypothesize that cultural fine-tuning operates under a representational bottleneck: instead of freely expanding the model's capacity to encode every cultural distinction, fine-tuning collapses cultural representations into a narrow activation subspace. To test this hypothesis, we probe distributed representations over feed-forward network neurons and compare the activated neuron sets before and after fine-tuning. Our analysis shows that fine-tuning uses a largely decoupled but restricted activation space, limiting the model's ability to preserve heterogeneous cultural profiles. This provides evidence that cultural flattening is not merely a behavioral artifact but linked to how cultural information is reorganized during optimization (§\ref{sec:mechanistic-analysis}).
}

\rw{
Our main contributions are threefold:
\begin{itemize}
  \item We introduce a unified evaluation framework that treats cultural alignment and cultural diversity as complementary axes, enabling joint measurement of whether models match human values while preserving cross-cultural variation.
  \item Applying this framework to six LLMs, we show that alignment gains consistently coincide with reduced behavioral diversity.
  \item We provide a mechanistic account of cultural flattening grounded in the low-rank simplicity bias, showing that fine-tuning compresses diverse cultural values into a low-rank subspace and thereby marginalizes minority cultures.
\end{itemize}
}

\section{Related Work}

\paragraph{Cultural Alignment.} With the widespread deployment and application of LLMs globally, imbuing these models with robust cultural awareness has become crucial. To achieve this, prior research has explored various optimization strategies, ranging from lightweight in-context learning \cite{AlKhamissi2024InvestigatingCA, choenni2025selfalignmentimprovingalignmentcultural, ki2025multiplellmagentsdebate, pham-etal-2025-cultureinstruct, shetty2025vital} and supervised fine-tuning \citep{huang2024acegpt, NEURIPS2024_9a16935b, li2024culturepark, Xu2024SelfPluralisingCA, jiang-etal-2025-language} to reinforcement learning preference optimization \cite{feng2025culfitfinegrainedculturalawarellm, zhao-culturerl-internal, yuan2026culturalpalettepluralisingculture}. Despite these promising advances, current evaluation and optimization paradigms remain largely centered on alignment-centric metrics that take human labels as references and quantify distances between model outputs and those references, a focus that leaves post-alignment behavioral diversity underexamined and may obscure underlying representational collapse. This raises a concern: because aggregate alignment scores can disproportionately reflect the dominant human preference, a model may score favorably by anchoring its behavior to that preference, even as its response diversity collapses. Moving beyond this one-dimensional view, we shift the reference point from human responses to the model itself, formalizing discrepancies across its behaviors as diversity and thereby enabling a systematic analysis of the alignment--diversity interplay.

\paragraph{Cultural Diversity.} Maintaining cultural and behavioral diversity is essential for LLMs to successfully integrate into and serve multifaceted human societies. To this end, several recent works have investigated whether deployed models can faithfully reproduce the rich diversity of human conceptual and behavioral distributions \cite{santurkar2023opinionslanguagemodelsreflect, hu-collier-2024-quantifying, Schrder2025LargeLM, wang2025largelanguagemodelsreplace}. Their findings consistently suggest that such socio-cultural diversity is difficult to capture accurately; crucially, recent studies have also begun to note that alignment-oriented optimization may unintentionally suppress behavioral diversity in models \citep{wu2024generativemonoculturelargelanguage, murthy-etal-2025-one}, raising critical concerns about post-alignment behavioral homogenization. In stark contrast to these existing studies that remain confined to post-hoc, black-box observations of diversity degradation, we venture into the post-training optimization process itself to provide structural insights. Specifically, we scrutinize the dynamic trajectories of multicultural fine-tuning to address two fundamental questions: (i) how superficial gains in alignment scores mask a latent collapse in representational and behavioral diversity; and (ii) whether this alignment--diversity trade-off is merely a superficial behavioral anomaly or an intrinsic, structural consequence of neural network optimization.

\section{Measuring the Alignment–Diversity Trade–off}\label{sec:preliminaries}
In this work, we propose a framework that measures both the benefits of cultural alignment and the costs of losing cultural diversity.

\paragraph{Tasks.} To formalize this, we first delineate how cultural awareness is evaluated: grounded in the sociological premise that culture manifests as structured patterns of behavior \cite{kroeber1952culture}, such evaluations conventionally measure the behavioral congruence between prompted model simulations and target human demographics \cite{joshi2024personas}. Formally, given a set of cultural contexts $C$ and a questionnaire set $Q$, we evaluate the model $\mathbb{M}$ through \textit{role-play prompting}. For each cultural context $c \in C$, the model is conditioned on culture-specific personas $p \sim P_c$ (e.g., "Imagine you are a Chinese female teacher") and asked to respond to survey questions $q \in Q$. This setup enables the model’s culturally conditioned behavioral patterns to be elicited and compared against human reference populations.

\paragraph{Alignment.}
Under this framework, \textit{Cultural Alignment} measures whether the model can faithfully reproduce the behavioral tendencies of the target sociodemographic group it is conditioned to simulate. Formally, we define the alignment score of model $\mathbb{M}$ as:
\begin{equation}
\mathcal{A}(\mathbb{M})=\mathbb{E}_{q,c,p}
\left[\mathcal{S}
\big(\mathbb{M}(q|p,c), \mathbb{H}(q|p,c)\big)
\right]
\label{eq:alignment}
\end{equation}

where $\mathbb{M}(q|p,c)$ denotes the model response under persona $(p,c)$, $\mathbb{H}(q|p,c)$ denotes the corresponding human behavioral reference, and $\mathcal{S}(\cdot,\cdot)$ measures behavioral similarity. A higher $\mathcal{A}(\mathbb{M})$ indicates stronger behavioral consistency with the target cultural group.

\paragraph{Diversity.}
While alignment evaluates behavioral fidelity within each cultural subgroup, \textit{Cultural Diversity} evaluates whether responses remain behaviorally distinguishable across different cultural subgroups. Specifically, a diverse model should preserve localized behavioral variation instead of collapsing toward a homogeneous response distribution. We therefore define the diversity score as:
\begin{equation}
\mathcal{D}(\mathbb{M})
=
\mathbb{E}_{q,p_i,p_j}
\left[
1 -
\mathcal{S}
\big(
\mathbb{M}(q|p_i,c_i),
\mathbb{M}(q|p_j,c_j)
\big)
\right]
\label{eq:diversity}
\end{equation}

where $(p_i,c_i)$ and $(p_j,c_j)$ are personas sampled from different cultural groups ($c_i \neq c_j$). Crucially, distinct from the monotonic optimization of $\mathcal{A}(\mathbb{M})$ where higher scores strictly denote better fidelity, higher diversity alone does not necessarily imply better cultural realism; meaningful diversity should remain grounded in human behavioral distributions. At the other extreme, however, excessive reductions in diversity signal cultural flattening, whereby culturally distinct behaviors collapse into increasingly homogeneous outputs.

Importantly, $\mathcal{A}(\mathbb{M})$ and $\mathcal{D}(\mathbb{M})$ are not independent objectives, but two complementary perspectives derived from the same underlying similarity function $\mathcal{S}(\cdot,\cdot)$. This shared formulation provides a unified basis for comparing behavioral fidelity and cross-cultural differentiation within a common evaluative space. In §~\ref{sec:experiment-setup}, we further instantiate $\mathcal{S}(\cdot,\cdot)$ using \textit{Soft Accuracy} and operationalize both objectives under the same measurement framework.

\section{Empirical Evidence of the Alignment–Diversity Trade-off}\label{sec:experiment-on-cultural-alignment}

In this section, we empirically assess the alignment--diversity trade-off. We first introduce the evaluation setup (§\ref{sec:experiment-setup}) and then present the observed trade-off between cultural alignment and diversity (§\ref{sec:trade-off-overview}). Next, we investigate the behavioral mechanisms driving these alignment gains at the expense of diversity (§\ref{sec:mech-pseudo-alignment}), ultimately unpacking the severe marginalization risks (§\ref{sec:case-risk}).

\begin{table*}[ht]
\centering
\small
\setlength{\tabcolsep}{3pt}
\resizebox{\textwidth}{!}{
\begin{tabular}{lcccccccccccccccccc}
\toprule
 & \multicolumn{3}{c}{\textbf{Gemma-2}} & \multicolumn{3}{c}{\textbf{Llama-3}} & \multicolumn{3}{c}{\textbf{Llama-3.1}} & \multicolumn{3}{c}{\textbf{Qwen-2.5}} & \multicolumn{3}{c}{\textbf{Qwen-3}} & \multicolumn{3}{c}{\textbf{Mistral}} \\
\cmidrule(lr){2-4} \cmidrule(lr){5-7} \cmidrule(lr){8-10} \cmidrule(lr){11-13} \cmidrule(lr){14-16} \cmidrule(lr){17-19}
 & W/O & SFT & SFT-E & W/O & SFT & SFT-E & W/O & SFT & SFT-E & W/O & SFT & SFT-E & W/O & SFT & SFT-E & W/O & SFT & SFT-E \\
\midrule
\textbf{Align} & 67.22 & \redbg{5}68.93 & \redbg{8}69.42 & 63.32 & \redbg{20}69.79 & \redbg{15}68.04 & 62.69 & \redbg{25}70.23 & \redbg{20}68.72 & 61.29 & \redbg{25}69.67 & \redbg{25}69.67 & 63.48 & \redbg{10}66.67 & \redbg{5}65.39 & 58.11 & \redbg{40}69.99 & \redbg{40}69.89 \\
\textbf{Div} & 17.52 & \bluebg{30}5.68 & \bluebg{60}0.52 & 21.58 & \bluebg{30}9.95 & \bluebg{30}10.58 & 24.48 & \bluebg{50}3.91 & \bluebg{40}5.22 & 31.18 & \bluebg{50}5.08 & \bluebg{60}3.39 & 21.26 & \bluebg{30}9.98 & \bluebg{30}7.15 & 26.92 & \bluebg{50}5.70 & \bluebg{50}4.96 \\
\bottomrule
\end{tabular}
}
\caption{Model alignment and diversity performance comparison. Align: Alignment, Div: Diversity. Cell background color intensity indicates the magnitude of change relative to the Base model; red indicates an increase, and blue indicates a decrease \textbf{[Human Reference: Align=100, Div=37.9]}.}
\label{tab:performance}
\end{table*}

\subsection{Experiment Setup}\label{sec:experiment-setup}

\paragraph{Evaluation Benchmark.} Our evaluation is built upon the World Values Survey 7 (WVS-7) \cite{Haerpfer2022} providing a globally representative sample of social and political values, administered in local languages to ensure cultural fidelity. To capture a broad cultural spectrum, we selected 14 countries across various continents: NGA, MAR, TUN, USA, BOL, URY, CHN, TJK, JOR, RUS, SRB, CYP, AUS, and NZL. For each country, the dataset includes 80 individual respondents, with each respondent covering responses to 31 survey items across representative value topics, where each item is paraphrased into three distinct variants to obtain multiple responses. Detailed descriptions and statistics of the WVS, along with a comprehensive mapping between country abbreviations and their full names, are presented in Appendix \ref{appendix:wvs}.

\paragraph{Model Selection.} Our evaluated models encompass six mainstream LLMs, spanning four model families with variations in both architectural versions and parameter scales: Llama-3-8B-Instruct, Llama-3.1-8B-Instruct \cite{Touvron2023LLaMAOA}, Qwen2.5-7B-Instruct \cite{Bai2023QwenTR}, Qwen3-4B \cite{yang2025qwen3technicalreport}, gemma-2-9B-it \cite{Riviere2024Gemma2I}, and Mistral-7b-Instruct-v0.3 \cite{Jiang2023Mistral7}.

\paragraph{Prompt Design.} We instantiate simulation respondents through persona prompts that integrate macro-level national identities with fine-grained demographic attributes, including \textit{country, region, sex, age, social class, education level, and marital status} \citep{AlKhamissi2024InvestigatingCA}. All prompts are formulated in English as a unified baseline, which helps isolate cultural alignment from the confounding effects of varying multilingual proficiencies across models \citep{NEURIPS2024_9a16935b, Xu2024SelfPluralisingCA}. Detailed prompt templates are provided in Appendix \ref{appendix:prompt-design}.

\paragraph{Fine-tuning Methods.} Our training corpus is also derived from the WVS-7 to ensure an in-domain evaluation, while maintaining a strict held-out separation between the training and evaluation sets. Specifically, the training data is curated by sampling 20 unique respondents per country and their responses to these aforementioned topics; in this process, we explicitly ensure the diversity of responses to prevent the training set from falling into a single pattern. Based on these data settings, we train two distinct models via \textit{supervised fine-tuning}, where both models are fine-tuned for a single epoch under identical hyperparameter configurations. We denote the model trained on the full dataset as \textit{SFT}, and the variant trained on the corpus \textit{\underline{E}}xcluding the United States and Australia as \textit{SFT-E}, aiming to evaluate mainstream cultural influences. We provide the training corpus statistics in Appendix \ref{appendix:wvs-trainingset}, with detailed fine-tuning implementation described in Appendix \ref{appendix:detail-sft-implementation}.

\paragraph{Similarity Metric.}
To operationalize the similarity function $\mathcal{S}(\cdot,\cdot)$ in Eqs.~\ref{eq:alignment} and \ref{eq:diversity}, we instantiate it using \textit{Soft Accuracy (SA)}. Unlike exact-match accuracy, SA assigns partial credit to ordinal-scale responses, thereby better capturing graded behavioral similarity across culturally conditioned outputs.

Formally, for a pair of responses $(y_i, y_j)$ to question $q$, SA is defined as:
\begin{equation}
SA(y_i, y_j) =
\begin{cases}
1 - \dfrac{|y_i - y_j|}{|q| - 1} & \text{if } \text{IsOrd}(q), \\
\mathbb{I}(y_i = y_j) & \text{otherwise},
\end{cases}
\end{equation}

where $\text{IsOrd}(q)$ indicates whether $q$ is an ordinal-scale item, and $|q|$ denotes the number of response options.

Within our framework, the same SA metric serves two complementary roles. In Eq.~\ref{eq:alignment}, it measures behavioral consistency between model outputs and human references. In Eq.~\ref{eq:diversity}, it measures behavioral similarity across culturally distinct personas, where lower similarity indicates stronger preservation of culturally localized behaviors. However, diversity should remain grounded in realistic human behavioral distributions, as excessively high inter-group similarity may lead to cultural flattening.

\subsection{Trade-off Overview}\label{sec:trade-off-overview}

Table \ref{tab:performance} exhibits the detailed alignment and diversity scores across various base models and fine-tuning strategies. Along the \textit{Align} row, cultural fine-tuning consistently enhances alignment across all architectures, yielding gains ranging from 1.7 (Gemma-2) to 11.9 (Mistral). This pervasive prevalence of red-shaded cells across the alignment metrics serves as a visual indicator that cultural fine-tuning reliably optimizes models toward higher behavioral consistency with target cultural groups.

\textit{However, these alignment gains are accompanied by a substantial reduction in cultural diversity, revealing \textbf{a systematic trade-off between behavioral fidelity and cultural variation.}} Along the \textit{Div} row, diversity scores for all fine-tuned models collapse below 11.0. This widespread plunge leaves a pronounced gap against the empirical human reference of 37.9 and fails to reach even the lower bound of the vanilla baselines (which uniformly stay above 17.5), {statistically implying an over 90\% probability that the fine-tuned cohort will generate identical responses even under substantially different cultural contexts.} The trade-off is particularly severe in Gemma-2, where a modest alignment gain of 2.2 coincides with a dramatic 17-point reduction in diversity, resulting in a near-zero diversity score of 0.5.

Such extreme behavioral convergence suggests that fine-tuning may improve alignment not through nuanced cultural understanding, but by collapsing culturally diverse behaviors toward a dominant response pattern.

\paragraph{Scope of Discussion.}
To better isolate the origin of this behavioral convergence, we characterize the observed alignment--diversity trade-off primarily as a \textit{fine-tuning-induced dynamic} along each model's tuning trajectory, rather than as a consistently static cross-model relationship. Across architectures, no significant correlation is observed within either the SFT or SFT-E cohort, nor when the two fine-tuned cohorts are pooled ($p>0.05$). Instead, the trade-off manifests consistently through the within-model shifts induced by fine-tuning, where alignment gains coincide with substantial losses in behavioral diversity (Appendix~\ref{appendix:adcor}).

\subsection{Behavioral Dynamics of Trade-off}\label{sec:mech-pseudo-alignment}

We trace the behavioral evolution from base to fine-tuned models against human benchmarks to uncover the underlying dynamics of this optimization trade-off.

\begin{figure}[t]
\centering
  \includegraphics[width=1.\linewidth]{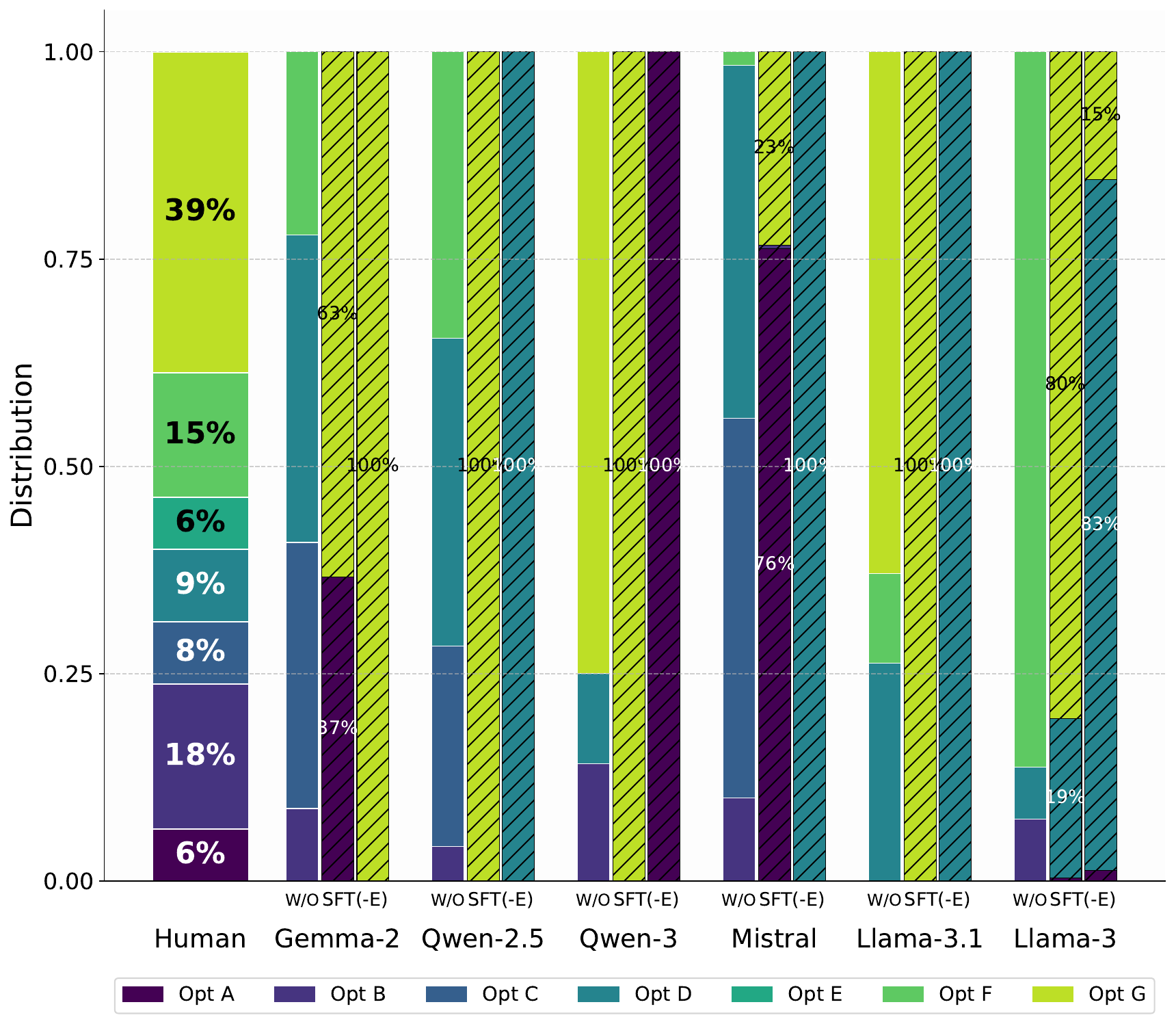}
    \caption{Behavioral response distributions across Human, Base (W/O), and fine-tuned configurations, where \textbf{hatched} bars denote \textbf{fine-tuned} models (Left: SFT; Right: SFT-E).}
\label{fig:notscale-analysis}
\end{figure}

To provide an intuitive understanding of these dynamics, we anchor our initial analysis on a granular visualization of the questionnaire item "How often do you attend religious services". This specific item is strategically selected because its highly heterogeneous human baseline spans the maximum spectrum of seven distinct choice categories, thereby offering a highly sensitive diagnostic baseline for tracking representational erosion (Figure \ref{fig:notscale-analysis}).

Mapping the models' post-fine-tuning response landscape against this rich baseline reveals a clear optimization shortcut: the fine-tuned outputs heavily anchor to the majority Option G, which represents the human plurality choice at a 39\% share. Notably, this behavioral pattern enforces a single stereotypical template that risks misapplying fixed values to cultural contexts where they are no longer mainstream—as evidenced by Qwen-3 fine-tuned under SFT-E, which stubbornly over-amplifies Option A into the dominant response even in contexts where it is a minority preference endorsed by only 6\% of humans.

An analysis of this case suggests the underlying nature of such alignment gains: conflating the emulation of dominant mainstream values with genuine cultural alignment, fine-tuned models suffer from a monolithic collapse that exclusively \textit{anchors} their outputs to a \textit{majority consensus}, thereby generating a form of trade-off.

\paragraph{Majority Anchoring Evaluation.} To provide complementary evidence for the observed diversity collapse and diagnose its underlying majority-anchoring behavior, we compute two simple diagnostic statistics across the evaluation set: (i) \textit{Majority Choice Rate}, the percentage of model responses that align with the most common human choice, measuring the inclination to follow plurality consensus; and (ii) \textit{Choice Pool Size}, the total count of unique options selected by the model, which captures the extent of behavioral homogenization. Results are summarized in Table \ref{tab:categorical-item}:
\begin{itemize}
  \item \textit{Majority Choice Rate}: Compared to vanilla baselines (W/O), both SFT and SFT-E models exhibit a continuous increase in this metric, demonstrating effective alignment with the plurality consensus; however, they unexpectedly surpass the human reference line (49.98\%), revealing a severe skewness toward dominant options.
  \item \textit{Choice Pool Size}: Even without fine-tuning, W/O models already exhibit a limited capacity to capture cultural diversity, scoring above 2.0 yet remaining far below the human reference of 3.61; such a limitation is further magnified under SFT and SFT-E regimes, where the metric collapses close to 1.0.
\end{itemize}

\begin{table}[t]
\centering
\tiny
\begin{tabular}{lcccccc}
    \toprule
    & \multicolumn{3}{c}{Majority Choice Rate (\%)} & \multicolumn{3}{c}{Choice Pool Size} \\
    \cmidrule(lr){2-4} \cmidrule(lr){5-7}
    Model & W/O & SFT & SFT-E & W/O & SFT & SFT-E \\
    \midrule
    Gemma-2-9B-it & 40.93 & 47.59 & 47.50 & 2.48 & 1.07 & 1.09 \\
    Qwen2.5-7B-Instruct & 40.59 & 55.59 & 52.14 & 2.66 & \textbf{1.06} & 1.23 \\
    Qwen3-4B & 38.53 & 52.29 & 45.78 & 2.39 & 1.42 & 1.45 \\
    Mistral-7b-Instruct-v0.3 & 40.16 & 54.80 & 51.51 & 2.64 & 1.19 & 1.31 \\
    Llama-3.1-8B-Instruct & 41.57 & 56.50 & 50.92 & 2.31 & 1.65 & 1.71 \\
    Llama-3-8B-Instruct & 40.76 & \textbf{57.74} & 51.49 & 2.06 & 1.51 & 1.41 \\
    \midrule
    \textit{Human Reference} &  & 49.98 &  &  & {3.61} &  \\
    \bottomrule
\end{tabular}
\caption{The majority choice rate (\%) and choice pool size across evaluated models on items. \textbf{Bold} values denote the global maximum of majority choice rate and the global minimum of choice pool size across all results.}
\label{tab:categorical-item}
\end{table}

By intertwining a heightened preference for plurality consensus with a drastic reduction in option variety, these two metrics collectively demonstrate that \textit{cultural fine-tuning mere \textbf{simulates consensus through behavioral homogenization}, thereby solidifying the empirical evidence for trade-off.}

 \begin{figure}[t]
\centering
  \includegraphics[width=1.\linewidth]{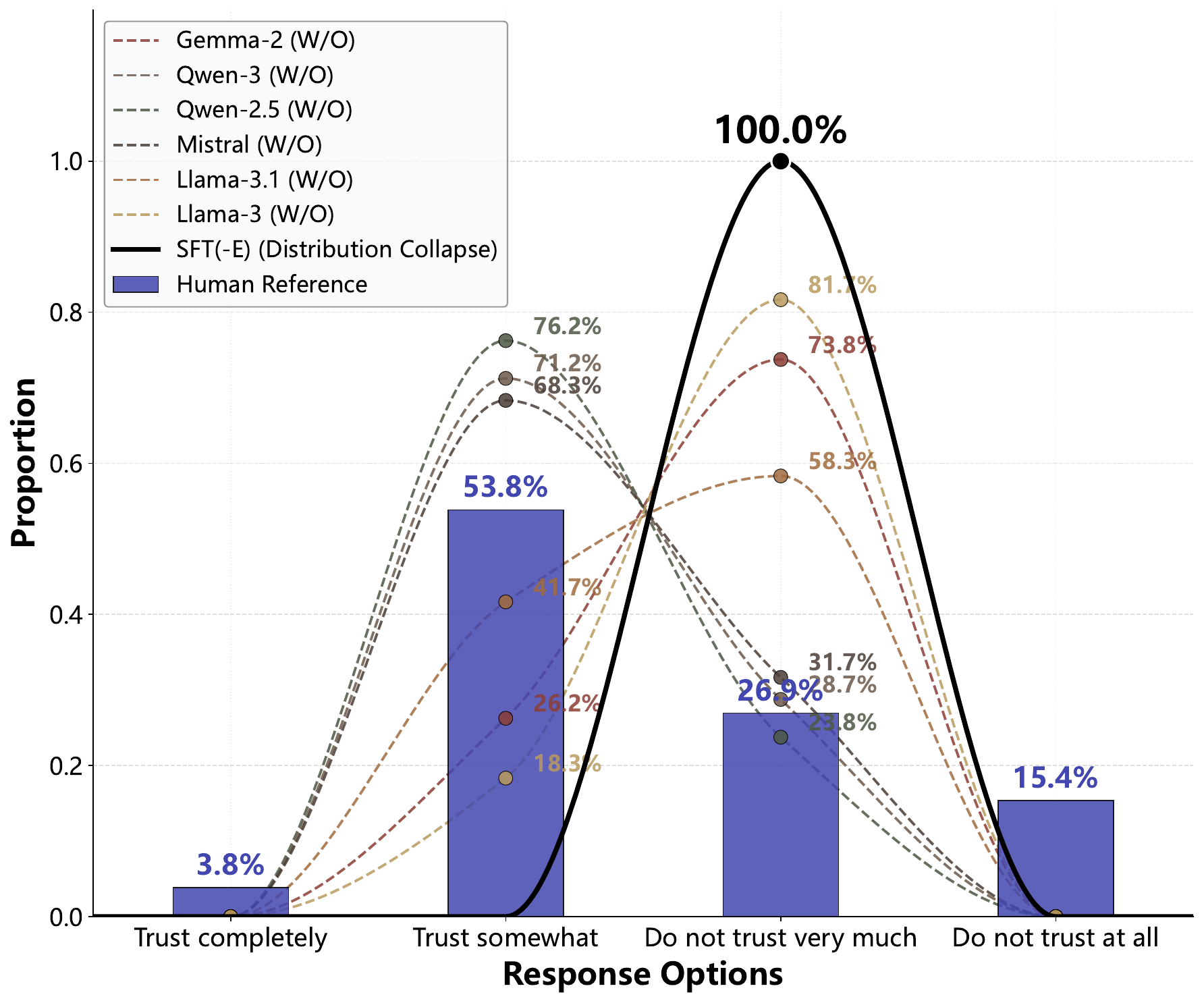}
    \caption{\textbf{Ordinal-item option distributions}. Bars indicate the Human Reference distribution. Scatter points show option proportions across settings, where dashed lines and the solid line represent the W/O and SFT(-E) settings respectively, connected via smooth trendlines.}
\label{fig:scale-analysis}
\end{figure}

\subsection{Bias and Risk Discussion}\label{sec:case-risk}

Moving beyond this trade-off, we shift from analyzing the models' learning dynamics to examining their unintended biases and associated risks.

\paragraph{Centralization Bias in Ordinal Items.} Intriguingly, when transitioning to ordinal items, the fine-tuning optimization does not always dictate a strict alignment with the absolute human plurality. Instead, we unexpectedly observe a strategic shift where the model retreats toward a moderately conservative stance. For instance, when evaluated on the item "How much do you trust people of another nationality?", while the human majority leans toward "Trust somewhat" ($53.8\%$), nearly all of the fine-tuned models undergo complete collapse, each concentrating its responses entirely ($100.0\%$) on "Do not trust very much." Notably, this choice is not arbitrary; rather, it strategically covers the broader human opinion spectrum within its immediate neighborhood (53.8\%, 26.9\%, and 15.4\%), as evidenced by Figure \ref{fig:scale-analysis}.

\paragraph{Marginalization Risks for Minority Cohorts.} Importantly, such monolithic response patterns directly expose historically under-represented cohorts to profound marginalization risks, a systemic failure for which we provide concrete empirical cases. Specifically, when evaluating a specific persona within the Nigerian context, fine-tuning severely disrupts the pre-trained alignment across all model families; this degradation is most notably exemplified by Qwen series, which experiences a drastic alignment drop from 51.52 to 37.59, with a relative decrease of 27.04\% (Appendix \ref{appendix:margin-risks}).

Although some cohort-specific degradation may be unavoidable during alignment, the systematic marginalization of the same cohort across model families warrants particular attention.

\paragraph{To sum up,} these combined empirical findings collectively corroborate that the superficial performance gains from fine-tuning are achieved through a straightforward optimization: \textit{the model captures the most prominent human values at \textbf{the cost of diversity}; critically, this shift inadvertently induces \textbf{cultural flattening}, thereby posing a severe risk of marginalizing under-represented minority cohorts.}

\begin{figure*}[t]
  \includegraphics[width=0.5\linewidth]{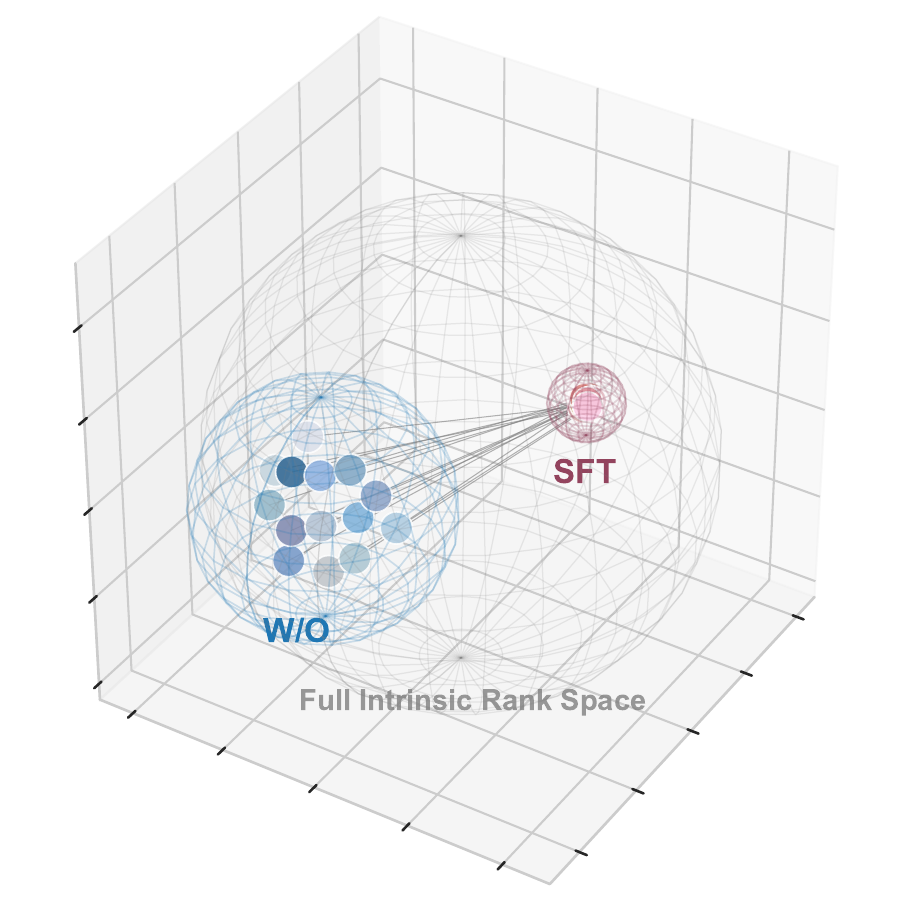}
  \includegraphics[width=0.5\linewidth]{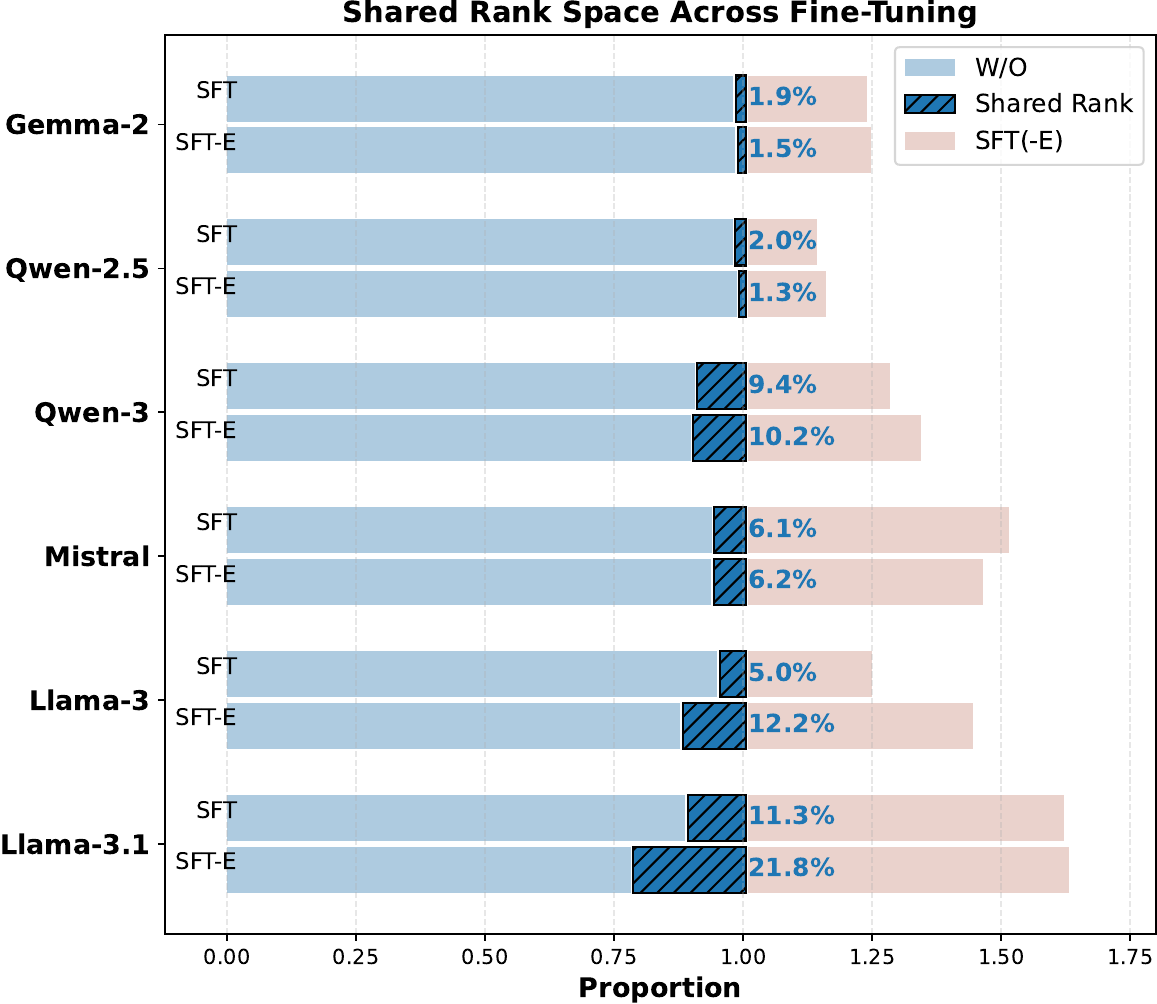}
  \caption{\textbf{Rank space distribution across tuning settings.} The \textbf{left} panel shows the projection of activation representation similarity for both W/O and SFT into a shared 3D space via Multidimensional Scaling; the \textbf{right} panel displays the shared rank space across fine-tuning for evaluated models, where the bar charts with \textbf{diagonal hatching} denote the proportion of \textbf{shared} activation ranks across fine-tuning.}
  \label{fig:low-rank-space}
\end{figure*}

\section{Mechanistic Analysis}\label{sec:mechanistic-analysis}

In this section, we present a mechanistic analysis of the alignment--diversity trade-off. We first propose our theoretical hypothesis alongside a dedicated probing method (§\ref{sec:hypothesis}), and then empirically assess evidence for the hypothesized mechanism (§\ref{sec:empirical-mechanics}).

\subsection{Hypothesis and Probing Method}\label{sec:hypothesis}

\paragraph{Hypothesis.} We hypothesize that this trade-off is driven by the \textit{Low-Rank Simplicity Bias} during neural network optimization, rather than being a mere superficial behavioral anomaly~\citep{huh2023lowranksimplicitybiasdeep, súkeník2024neuralcollapseversuslowrank, galanti2024sgdweightdecaysecretly}. Specifically, long-term convergence on massive data during pre-training establishes a stable, low-rank upper bound for the model's representations. Consequently, during fine-tuning, gradient updates are not globally flexible but are strictly confined to a narrow, leftover subspace. This restricted optimization space limits the model's representational capacity, forcing it to discard less salient cultural information and making it more susceptible to losses in behavioral diversity. We provide additional theoretical details and discussions regarding the hypothesis in Appendix \ref{appendix:low-rank-simp-bias-resource-allocation}.

\paragraph{Probing Method.} To parameterize the compressed rank space, we look into the activation space of \textit{Feed-Forward Network (FFN) neurons}. Existing literature views FFN neurons as key-value memories that store and project conceptual knowledge \citep{geva2021transformerfeedforwardlayerskeyvalue, Tang2024LanguageSpecificNT,Ying2025DisentanglingLA}. Accordingly, we use the activation patterns of FFN neurons as a measurable proxy for the effective low-rank structure of the model's cultural representations.

Formally, for a decoder-only Transformer at layer $l$, the FFN processing hidden state $h^l$ is defined as:
\begin{equation}
\text{FFN}^l(h^l)=W^l_\text{down} \cdot \theta(W^l_\text{up} \cdot h^l)
\end{equation}
where $W^l_\text{up}$ and $W^l_\text{down}$ are weight matrices and $\theta(\cdot)$ is activation function. We denote the $i$-th element of $\theta(W^l_\text{up} \cdot h^l)$ as the $i$-th neuron $n_{(l,i)}$ in layer $l$.

During inference, we assume that neurons with larger activation values contribute more strongly to culture-related processing and are therefore more likely to carry relevant cultural information. Based on this assumption, for each question $q \in Q$, we identify the activated-neuron set $\mathcal{N}_q$ using a question- and layer-specific threshold:
\begin{equation}
\mathcal{N}_q
=
\left\{
n_{(l,i)}
\mid
a_{(q,l,i)} \geq \mu{(q,l)}
\right\},
\label{eq:neuron-extract}
\end{equation}
where $a_{(q,l,i)}$ is the activation of neuron $n_{(l,i)}$ elicited by question $q$, and $\mu{(q,l)}$ is the $K$-th largest activation in layer $l$ for that question, with $K=100$. We then aggregate the activated neurons across all questions: $\mathcal{N}_{\mathbb{M}} = \bigcup_{q \in Q} \mathcal{N}_q$.

The cardinality of $\mathcal{N}_{\mathbb{M}}$ serves as a proxy for the breadth of the model's effective cultural representational space. A smaller $|\mathcal{N}_{\mathbb{M}}|$ indicates that the model repeatedly relies on a narrower set of FFN neurons, suggesting a more restricted representational space.

\subsection{Empirical Analysis}\label{sec:empirical-mechanics}

Figure~\ref{fig:low-rank-space} (\textit{left}-panel) illustrates the rank-space distribution of cultural representations across countries, comparing models without (W/O) and with (SFT) cultural fine-tuning. Each scatter point represents the cultural representation of a distinct country, and the Euclidean distances between points, estimated from activated-neuron-set overlap, measure cross-country representational divergence under each model setting. The blue and red wireframe spheres correspond to W/O and SFT, respectively, with their radii indicating the size of the effective rank space.

\paragraph{Spatial Decoupling.} The rank spaces of the W/O and SFT models exhibit a high degree of geometric decoupling. Specifically, within the full intrinsic rank space of the model, which is represented by the outer grey hypersphere, the cultural representational subspaces for the W/O and SFT conditions show minimal geometric overlap. This spatial segregation implies that cultural fine-tuning operates under a strict representational bottleneck, forced to optimize within a highly compact and independent subspace. Such structural restriction not only limits the capacity of the model to encode diverse cultural profiles but also introduces the risk of erasing valuable pre-trained cultural knowledge.

To quantify the extent to which the fine-tuned representational space departs from the W/O baseline, we calculate the \textit{baseline rank-space retention ratio}, which measures the proportion of W/O-activated neurons retained after fine-tuning: $|N_{\mathbb{M}}^{\text{(W/O)}} \cap N_{\mathbb{M}}^{\text{(SFT(-E))}}| / |N_{\mathbb{M}}^{\text{(W/O)}}|$.

Figure \ref{fig:low-rank-space} (\textit{right}-panel) presents this metric, with the empirical allocations visually highlighted by the hatched bars. We consistently observe that the shared activated-neuron set accounts for less than 10\% of the W/O space across most evaluation settings. This limited overlap suggests that cultural fine-tuning does not enrich the model's expressive capacity but instead appears to collapse existing cultural representations into a narrow, largely distinct activation subspace. Moreover, this representational shift is largely insensitive to the training corpus, as indicated by the comparable retention ratios under SFT and SFT-E.

\paragraph{Spatial Restriction.} Even though optimization occurs in a highly decoupled space, we observe that the model's total effective rank space does not undergo a corresponding expansion; instead, this process is confined to a profoundly constricted subspace. For most models, the incremental space accounts for less than 50\% of the W/O dimensions, dropping to a striking 25\% in Qwen-2.5.

In summary, these results are consistent with our hypothesis that cultural fine-tuning operates under a restrictive rank bottleneck that limits the model's capacity to encode nuanced cultural profiles. This structural constraint offers a geometric perspective on the recurring reduction in cultural diversity observed throughout the alignment process.

To clarify, this paper focuses on characterizing the trade-off between alignment and diversity; the mechanistic analysis offers an interpretive lens on this observed behavior, rather than redirecting our primary scope toward theoretical neural network derivation or constituting a rigorous causal validation.

\section{Conclusion}
In this work, we first mathematically formalize cultural alignment and diversity within a unified evaluation framework, through which we uncover a systematic trade-off across LLMs: improvements in cultural alignment consistently come at a substantial cost to diversity, pushing models toward dominant majority preferences and monolithic response patterns that overlook the cultural perspectives of marginalized cohorts and ultimately lead to severe cultural flattening. Moving beyond post-hoc behavioral auditing, we further conduct a mechanistic analysis through the lens of the low-rank simplicity bias in neural network optimization. Our results provide empirical evidence that cultural fine-tuning operates within a compact, largely decoupled activation subspace, offering a structural perspective on the observed reduction in diversity. Together, these findings underscore the need to explore new strategies for multicultural alignment that improve cultural fidelity while preserving pluralistic representation.

\section*{Limitations}
We now discuss the limitations of our work:
\begin{itemize}[leftmargin=*, noitemsep]
\item \textbf{Limited Coverage of Alignment Strategies.}
Our empirical analysis is restricted to SFT. Although SFT remains a common approach to aligning LLMs, preference- and reinforcement-based methods, such as DPO, PPO, and GRPO, may exhibit different alignment--diversity dynamics. We do not systematically evaluate these methods because constructing reliable, culturally grounded preference annotations and reward signals remains particularly challenging.

\item \textbf{Limited Open-Ended and Dynamic Evaluation.}
Our diagnostic evaluation relies on static benchmarks, which cannot fully capture the context-dependent and evolving nature of cultural values. Future work should extend this evaluation to open-ended, interactive environments and investigate whether and how deployment-time interactions entrench biases within model representations over time.

\item \textbf{Lack of a Mitigation Strategy.}
This work does not develop a concrete strategy to mitigate cultural flattening. Similar representation collapse has also been observed in multimodal models, where knowledge distillation has been explored as a mitigation approach \citep{chaudhuri2025closer}. By analogy, future work could investigate culture-aware knowledge distillation to preserve information associated with both dominant and underrepresented cultures during alignment.

\end{itemize}

\bibliography{custom}

\appendix

\section{The Usage of AI}\label{sec:use-of-ai}
In this work, the application of AI is strictly limited to aiding and polishing academic writing, e.g., the description of evaluation framework, with no involvement in core research processes.

\section{Detailed Description and Statistical Analysis of the WVS} \label{appendix:wvs}
\subsection{Detailed Description of WVS}
The World Values Survey (WVS) collects responses to a range of questions broadly categorized into social, cultural, material, governmental, ethical, and economic domains, with data drawn from demographically controlled population samples worldwide \cite{Haerpfer2022}. Its seventh wave (WVS-7), conducted between 2017 and 2021, incorporates region-specific modules alongside globally standardized categories. Comprising 259 questions, WVS-7 is designed to include indicators aligned with multiple United Nations Sustainable Development Goals. Administered as a questionnaire, the survey targets selected samples from the general population, with all questions localized to the native or dominant regional languages.

The value topics used in this work are as follows:
(1) Friends, (2) People of a different race, (3) Immigrants/foreign workers, (4) Basic kinds of attitudes concerning society, (5) People of another religion, (6) People of another nationality, (7) Major Companies, (8) Private banks, (9) The United Nations (UN), (10) International Monetary Found (IMF), (11) The World Bank (WB), (12) The World Health Organization (WHO), (13) Increases the crime rate, (14) Increases the risks of terrorism, (15) Helps poor people establish new lives, (16) Losing my job or not finding a job, (17) Not being able to give one's children a good education, (18) Freedom and Equality - Which more important, (19) Freedom and security - Which more important, (20) How often do you attend religious services, (21) To make sense of life after death vs To make sense of life in this world, (22) Interest in politics, (23) Signing a petition, (24) Joining in boycotts, (25) local level, (26) Votes are counted fairly, (27) Election officials are fair, (28) How much would you say the political system in your country allows people like you to have a say in what the government does?, (29) Having a strong leader who does not have to bother with parliament and elections, (30) Having experts, not government, make decisions according to what they think is best for the country, (31) Having a system governed by religious law in which there are no political parties or elections.

Although the WVS-7 data was collected between 2017 and 2021 and may have certain temporal limitations, it remains one of the most recent and comprehensive datasets available for this research; furthermore, our sampling within each culture covers diverse demographic variables—including gender, age, and social class—to mitigate representation bias. The concerns regarding the values timeliness, contextual dynamics and sample representativeness reflect systemic challenges within the field, rather than being specific to this work. Constructing such corpora involves significant resource requirements and privacy considerations, which transcend the scope of this study.

\subsection{Abbreviation and Full Country Name}
\begin{table}[t]
\centering
\small
\begin{tabular}{ll}
\toprule
\textbf{Abbreviation} & \textbf{Full Country Name} \\
\midrule
AUS & Australia \\
BOL & Bolivia \\
CHN & China \\
CYP & Cyprus \\
JOR & Jordan \\
MAR & Morocco \\
NGA & Nigeria \\
NZL & New Zealand \\
RUS & Russia \\
SRB & Serbia \\
TJK & Tajikistan \\
TUN & Tunisia \\
URY & Uruguay \\
USA & United States \\
\bottomrule
\end{tabular}
\caption{Mapping of ISO three-letter country abbreviations to their full names.}
\label{tab:country_mapping}
\end{table}

To facilitate cross-referencing of the evaluated regional cohorts, the full country names corresponding to each ISO three-letter abbreviation are detailed in Table~\ref{tab:country_mapping}.

\section{Prompt Design}\label{appendix:prompt-design}
\begin{figure}[t]
\centering
  \includegraphics[width=1.\linewidth]{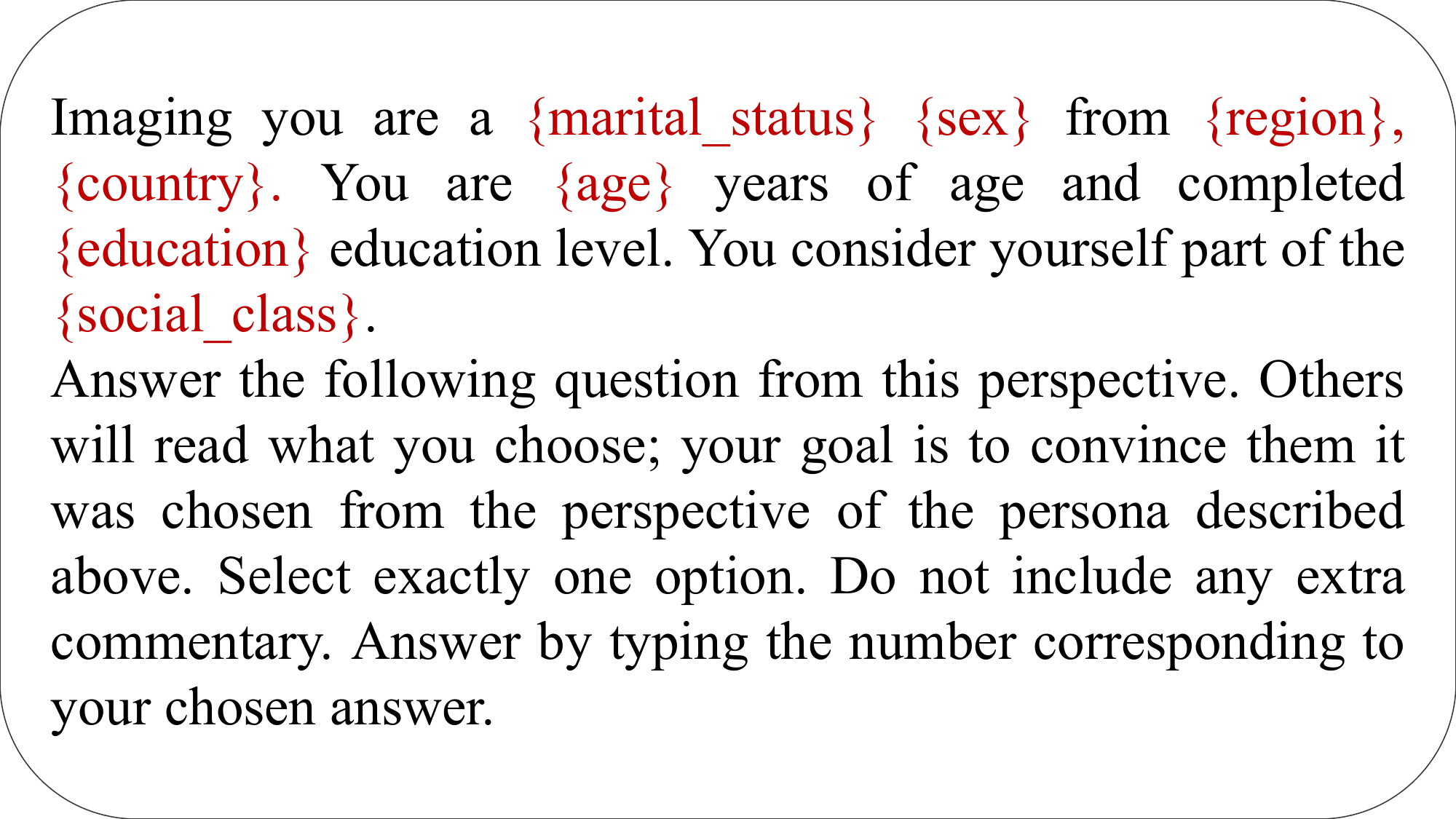}
    \caption{Persona Prompting used in this work.}
\label{fig:prompt}
\end{figure}

The instruction first initiates contextual persona inception by dynamically injecting comprehensive demographic profiles, including marital status, sex, geographic region, country, age, education level, and social class, which shifts the model's attention weights away from globally averaged pre-training patterns toward targeted, localized cultural priors, as described in Figure \ref{fig:prompt}.

Concurrently, the prompt applies strict Behavioral Conditioning by embedding a socially conditioned game-theoretic objective ("your goal is to convince them it was chosen from the perspective of the persona described above"), which successfully strips away the over-smoothed safety neutrality typical of conversational assistants.

Finally, through rigid Output Structure Formalization, the model is restricted to a single categorical choice token and explicitly barred from generating any extra commentary ("Select exactly one option. Do not include any extra commentary"), ensuring a clean and deterministic extraction of the model's underlying associative probability distributions while exposing the structural space collapse and marginalization risks induced by the fine-tuning process.

\section{Statistical Analysis of the Fine-tuning Corpus}\label{appendix:wvs-trainingset}

\subsection{Country-Level Distribution of Training Samples}
\begin{table}[t]
\small
\centering
\begin{tabular}{lc}
\toprule
\textbf{Country} & \textbf{\#Number} \\
\midrule
USA & 1842 \\
BOL & 1746 \\
URY & 1743 \\
CHN & 1410 \\
TJK & 1695 \\
JOR & 1662 \\
RUS & 1773 \\
SRB & 1830 \\
CYP & 1674 \\
NGA & 1788 \\
MAR & 1860 \\
TUN & 1779 \\
AUS & 1839 \\
NZL & 1683 \\
\midrule
SFT & 24324 \\
SFT-E & 20643 \\
\bottomrule
\end{tabular}
\caption{Distribution of the dataset across different countries and training configurations.}
\label{tab:data-distribution}
\end{table}

Table \ref{tab:data-distribution} provides an overview of the dataset composition across the fourteen target countries, as well as the total volume of samples used for the SFT and SFT-E training configurations. To ensure the quality and reliability of the training corpus, a rigorous data cleaning process was conducted. Specifically, entries containing null values, empty strings, or ambiguous labels such as "nonknown" were identified and filtered out prior to training. This preprocessing step ensures that the models are aligned using high-quality, meaningful cultural data, thereby enhancing the robustness of the resulting cultural language models.

\subsection{Distribution by Number of Answer Options}
\begin{table}[t]
\centering
\small
\begin{tabular}{crrrr}
\toprule
& \multicolumn{2}{c}{\textbf{SFT}}
& \multicolumn{2}{c}{\textbf{SFT-E}} \\
\cmidrule(lr){2-3}
\cmidrule(lr){4-5}
\textbf{\# Options}
& \textbf{Count} & \textbf{Perc. (\%)}
& \textbf{Count} & \textbf{Perc. (\%)} \\
\midrule
2        & 4,083  & 16.79  & 3,489  & 16.90 \\
3        & 4,953  & 20.36  & 4,236  & 20.52 \\
4        & 14,454 & 59.42  & 12,204 & 59.12 \\
$\geq 5$ & 834    & 3.43   & 714    & 3.46  \\
\midrule
\textbf{Total}
& \textbf{24,324} & \textbf{100.00}
& \textbf{20,643} & \textbf{100.00} \\
\bottomrule
\end{tabular}
\caption{Distribution of training samples by the number of answer options in SFT and SFT-E. SFT-E excludes samples from the United States and Australia.}
\label{tab:option-count-distribution}
\end{table}

Table~\ref{tab:option-count-distribution} shows that SFT and SFT-E have highly similar option-count distributions, with the proportion of each category differing by at most 0.31 percentage points. In both datasets, training samples are predominantly concentrated in questions with two, three, or four answer options, which together account for 96.57\% of SFT and 96.54\% of SFT-E.

\begin{table}[t]
\centering
\small
\resizebox{\columnwidth}{!}{\begin{tabular}{ccrrrr}
\toprule
\textbf{Dataset}
& \textbf{\# Opts}
& \textbf{Answer 1}
& \textbf{Answer 2}
& \textbf{Answer 3}
& \textbf{Answer 4} \\
\midrule
\multirow{3}{*}{\textbf{SFT}}
& 2 & 1,311 (32.11\%) & 2,772 (67.89\%) & -- & -- \\
& 3 & 1,155 (23.32\%) & 1,560 (31.50\%) & 2,238 (45.18\%) & -- \\
& 4 & 3,213 (22.23\%) & 4,119 (28.50\%) & 4,062 (28.10\%) & 3,060 (21.17\%) \\
\midrule
\multirow{3}{*}{\textbf{SFT-E}}
& 2 & 1,113 (31.90\%) & 2,376 (68.10\%) & -- & -- \\
& 3 & 972 (22.95\%) & 1,281 (30.24\%) & 1,983 (46.81\%) & -- \\
& 4 & 2,802 (22.96\%) & 3,435 (28.15\%) & 3,321 (27.21\%) & 2,646 (21.68\%) \\
\bottomrule
\end{tabular}}
\caption{Answer-label distributions for questions with two, three, and four valid answer options in SFT and SFT-E. Each cell reports the number of samples, with the corresponding percentage in parentheses. \textit{Don't know} options are excluded.}
\label{tab:answer-label-distribution}
\end{table}

\subsection{Answer-Label Distribution and Balance}
Table~\ref{tab:answer-label-distribution} shows that the training labels remain broadly distributed across the valid answer options rather than collapsing onto a single answer position. Although perfect uniformity is neither expected nor desirable because the labels preserve empirical response patterns from the WVS, no answer position consistently dominates across question types: Answer~2 is most frequent for two-option questions, Answer~3 for three-option questions, while four-option questions exhibit a relatively balanced distribution, with each answer accounting for approximately 21\%--29\% of the samples. Moreover, the highly similar distributions between SFT and SFT-E indicate that excluding the United States and Australia does not introduce a systematic label shift. Together, these observations alleviate concerns that the reported alignment--diversity patterns are artifacts of a globally dominant answer label or answer-position bias.

\section{Detailed SFT Implementation} \label{appendix:detail-sft-implementation}

\paragraph{Training Protocol and Hyperparameters} To understand the mechanistic alignment--diversity trade-off, we fine-tune the backbone models on downstream cultural instruction datasets using the \texttt{SFTTrainer} from the \texttt{trl} library. All models are optimized for one epoch with a linear learning-rate scheduler, and fine-tuning is conducted on two NVIDIA A100 GPUs.

\paragraph{Evaluation and Hardware Configuration}Crucially, to eliminate downstream generation stochasticity and guarantee strict empirical reproducibility, the text generation during the evaluation phase is executed using greedy decoding by anchoring the temperature parameter precisely to $\tau = 0$. Both the pipeline training and downstream diagnostic evaluations are implemented on a distributed hardware configuration consisting of four NVIDIA Tesla P100 GPUs, utilizing PyTorch and the Hugging Face \texttt{Transformers} library ecosystem.

\section{Alignment--Diversity Correlations Across Tuning Conditions}\label{appendix:adcor}

\begin{table}[t]
\centering
\small
\begin{tabular}{lr}
\toprule
\textbf{Condition} & \textbf{Pearson's $r$ ($p$-value)} \\
\midrule
W/O                     & $-0.909^{*}$ ($p=0.012$) \\
SFT                     & $-0.614$ ($p=0.195$) \\
SFT-E                   & $-0.536$ ($p=0.273$) \\
\textbf{W/O + SFT}               & $-0.957^{***}$ ($p<0.001$) \\
\textbf{W/O + SFT-E}             & $-0.928^{***}$ ($p<0.001$) \\
SFT + SFT-E             & $-0.471$ ($p=0.122$) \\
\textbf{W/O + SFT + SFT-E}       & $-0.927^{***}$ ($p<0.001$) \\
\bottomrule
\end{tabular}
\caption{Pearson correlations between alignment and diversity across model architectures under individual and pooled tuning conditions. For pooled conditions, observations from the corresponding states are concatenated. Statistical significance is denoted by $^{*}p<0.05$, $^{**}p<0.01$, and $^{***}p<0.001$.}
\label{tab:alignment-diversity-correlation}
\end{table}

Table~\ref{tab:alignment-diversity-correlation} reports the correlations between alignment and diversity across individual and pooled tuning conditions. Although the W/O cohort already exhibits a significant negative correlation, this relationship is not significant within either the SFT or SFT-E cohort, nor when the two fine-tuned cohorts are pooled. In contrast, more pronounced negative correlations emerge when W/O and fine-tuned states are considered jointly, highlighting the systematic alignment gains and diversity losses associated with fine-tuning. These results suggest that the alignment--diversity trade-off is not consistently expressed as a static cross-model property under a fixed tuning condition, but is primarily amplified along the behavioral transition induced by fine-tuning.

\section{Marginalization Risks for Minority Cohorts}\label{appendix:margin-risks}

\begin{table}[t]
\centering
\small
\setlength{\tabcolsep}{3pt}
\resizebox{\columnwidth}{!}{\begin{tabular}{lccc}
\toprule
\textbf{Model Name}
& \textbf{W/O}
& \textbf{SFT}
& \textbf{SFT-E} \\
\midrule
Gemma-2-9b-it
& 51.61
& 52.42\relup{1.57}
& 44.80\reldown{13.20} \\

Qwen2.5-7B-Instruct
& 51.52
& 37.59\reldown{27.04}
& 39.18\reldown{23.95} \\

Qwen3-4B
& 47.40
& 51.88\relup{9.45}
& 42.83\reldown{9.64} \\

Mistral-7B-v0.3
& 44.44
& 48.21\relup{8.48}
& 43.19\reldown{2.81} \\

Meta-Llama-3-8B
& 47.85
& 47.67\reldown{0.38}
& 42.92\reldown{10.30} \\

Meta-Llama-3.1-8B
& 50.36
& 42.47\reldown{15.67}
& 38.17\reldown{24.21} \\
\bottomrule
\end{tabular}}
\caption{Performance of different models on the lowest-aligned cohorts within the Nigerian population across tuning regimes. Scores are multiplied by 100. Smaller parenthetical values report relative changes from W/O, with green indicating an increase and red indicating a decrease.}
\label{tab:model-performance}
\end{table}

Table \ref{tab:model-performance} reveals a critical paradox: while fine-tuning regimes generally enhance the overall alignment performance of LLMs, they simultaneously expose a profound negligence toward specific minority cohorts. In the Nigerian context, the model’s ability to perceive and align with the most marginalized demographic—characterized by specific socioeconomic attributes—is significantly diminished post-fine-tuning. For instance, the drastic performance drop observed in the Qwen and Llama series after SFT-E highlights a systemic failure: the fine-tuning process, while optimizing for general population metrics, inadvertently erodes the model’s sensitivity to under-represented identities, thereby reinforcing the marginalization of these cohorts through reduced alignment granularity.

\section{Low-Rank Simplicity Bias in Cultural Alignment}\label{appendix:low-rank-simp-bias-resource-allocation}

The observed decline in cultural diversity during fine-tuning is rooted in the intrinsic low-rank simplicity bias of neural network optimization. During Supervised Fine-Tuning (SFT), gradient updates are not globally flexible; instead, they are strictly confined to a narrow, low-rank manifold established during the pre-training phase. This restriction creates a rank bottleneck that forces the model to prioritize a limited set of representational dimensions.

\paragraph{Polysemantic Competition and Interference.} Due to this restricted rank, neurons are forced to become polysemantic, encoding multiple cultural features within the same limited subspace. In our task, when the model optimizes for a specific cultural value (alignment), polysemantic collisions occur. Dominant alignment signals, which provide the steepest gradient for loss reduction, effectively interfere with and suppress more nuanced, marginalized cultural signals. This "interference" prevents the model from simultaneously representing the full spectrum of cultural heterogeneity

\paragraph{Rank Starvation and Geometric Collapse.} This phenomenon leads to a state of rank starvation, where the available degrees of freedom in the representational manifold are insufficient to accommodate high-dimensional cultural diversity. Optimization dynamics under weight decay and SGD favor solutions that collapse features toward their class means, a process known as Neural Collapse. Consequently, while the model successfully "aligns" to target cultural averages, it achieves this by physically "squeezing out" the variability of thought

To sum up, In the context of cultural SFT, the "Flattening" effect is not merely a data frequency issue but a physical consequence of resource allocation under rank constraints. The model essentially trades off its latent representational capacity for optimization efficiency, leading to a singular, low-rank representation that lacks the diverse "rank budget" necessary to simulate heterogeneous human populations.

\end{document}